-------------------------------------------------------------------------------

\documentstyle[12pt,aasms4]{article}

\def\gappr{\mathrel{\vcenter{\offinterlineskip \hbox{$>$}
    \kern 0.3ex \hbox{$\sim$}}}}
\def\lappr{\mathrel{\vcenter{\offinterlineskip \hbox{$<$}
    \kern 0.3ex \hbox{$\sim$}}}}

\received{}
\revised{}

\begin{document}

\setlength{\baselineskip}{24pt}

\title{3-D Simulations of Protostellar Jets in Stratified Ambient Media } 
\author{Elisabete M. de Gouveia Dal Pino } 
\affil{University of S\~ao Paulo, Instituto Astron\^omico e Geof\'isico\\
Av. Miguel St\'efano, 4200, S\~ao Paulo, SP 04301-901, Brazil 
\\ dalpino@astro1.iagusp.usp.br}\author{Mark Birkinshaw } 
\affil{Departmente of Physics \\
University of Bristol, Bristol, UK} 


\begin{abstract}

We present fully three-dimensional hydrodynamical simulations of radiative
cooling jets propagating into stratified isothermal ambient media with
power-law density and pressure distributions. The parameters used are mainly
suitable for protostellar jets but results  
applicable to extragalactic jets are also presented. Comparisons are made
with previous simulations of jets through homogeneous media. We find that
for radiative cooling jets propagating into regions where the ambient medium
has an increasing density (and pressure) gradient, the ambient gas tends to
compress the cold, low-pressure cocoon of shocked material that surrounds
the beam and destroy the bow shock-like structure at the head. The
compressing medium collimates the jet and promotes the development of
Kelvin-Helmholtz instabilities which cause beam focusing, wiggling and the
formation of internal traveling shocks, $close$ $to$ $the$ $head$, via pinching
along the beam. This remarkably resembles the structure of some observed
systems (e.g. Haro 6-5B northern and HH 24G jets). These effects are larger
for jets with smaller density ratio between jet and environment $\eta $
(tested for $\eta $=1, 3, and 10) and larger Mach number $M_a=v_j/c_a$
(tested for $M_a=$12 and 24, where $v_j$ is the jet velocity and $c_a$ the
ambient sound speed). 
In an ambient medium of decreasing density
(and pressure), the beam is poorly collimated  and relaxes, becoming faint.
This could explain ''invisible'' jet sections, like the gap between the
parent source and collimated beam (e.g., in HH30 jet). Although, on average,
jets propagating into an increasing (decreasing) density environment are
decelerated (accelerated) by the increasing (decreasing) ram pressure of the
ambient medium, we find that their propagation velocities have an
oscillating pattern. 
The internal traveling shocks that develop in jets propagating into positive
density gradient environments display a similar velocity variation, in
qualitative agreement with recent measurements of fluctuations in the
tangential velocity of the knots of Haro 6-5B jet. Finally, runs of
adiabatic jets into similar stratified environments indicate that they are
less affected by the effects of stratification than the cooling jets because
their higher pressure cocoons are better able to preserve the beam structure.
\end{abstract}


\clearpage

\section{Introduction}

Supersonic jets emanating from protostars and active galactic nuclei
propagate into complex media. Protostellar jets, for example, are immersed
in molecular clouds which are composed of many smaller dense clouds (e.g.
Mundt, Ray and Raga 1991, hereafter MRR; Bally and Devine 1994). 
Examining a sample of 15 protostellar jets, MRR have found that the flows 
are often poorly collimated close to the source and become better collimated
at large distances from their sources. The Haro 6-5B jet, in particular, 
after a section of strong expansion, narrows with increasing distance from the 
source, presumably because of reconfinement.
These observations  suggest that, in addition 
to some local collimation near the source, a very large scale collimation 
mechanism is at work, which is possibly due to the jet external 
environment (MRR). 

The question of what
happens to radio-emitting extragalactic jets propagating into a
non-homogeneous medium was discussed by 
Sanders (1983),
Falle and Wilson (1985a, b), 
Wiita, Rosen and Norman (1990) and
Hardee et al. (1992), who performed two-dimensional hydrodynamical
simulations of adiabatic, light jets. Since the radiative cooling distance
behind the post-shock gas in protostellar jets is smaller than the jet
radius, an adiabatic approach is inappropriate. Furthermore, the lack of
backflowing cocoons associated with protostellar jets, and the large advance
speeds of their working surfaces, suggest they are usually denser than their
surroundings. Earlier numerical modeling of jets emanating from protostars
were performed for beams propagating into homogeneous ambient media (e.g.,
Blodin, Fryxell and K\"onigl 1990, hereafter BFK; Gouveia Dal Pino and Benz
1993, hereafter GB93; Gouveia Dal Pino and Benz 1994, hereafter GB94;
Chernin et al. 1994, hereafter CMGB; Stone and Norman 1993 (hereafter SN),
1994a, b; Biro and Raga 1994). In this work, we examine the effects of
stratified environments on the structure of radiatively cooling and
adiabatic, initially
heavy jets with the aid of three-dimensional numerical simulations using the
SPH technique (e.g. GB93). Some of the results of these simulations and their
possible correlation with observed wiggling jets have been recently
summarized in Gouveia Dal Pino, Birkinshaw and Benz (1995, hereafter GBB).
We here describe in more detail the results for a range of environmental and
jet parameters. 

Previous work of jets propagating into homogeneous ambient media  has 
demonstrated that internal traveling
knots may be produced by small-period velocity variations in the ejection of
the jet 
(e.g., Raga et al. 1990, Raga and Kofman 1992, Kofman And Raga 1992, Hartigan
and Raymond 1992, GB94, SN).
This mechanism favors the formation of knots closer to the driving
source which become fainter and disappear at larger distances, as for
example, in jet systems like HH34 and HH111. In the present work, 
we find that the propagation of a radiative cooling jet into an atmosphere
of increasing density (pressure) may drive the formation of internal knots
closer to the jet head. This seems to be the case, for example, for the Haro
6-5B northern and HH 24G jets. 

The organization of this work is as follows. In \S 2, based on simple
one-dimensional arguments, we discuss the basic theoretical properties of
jets propagating into stratified ambient media and describe the assumptions
of our numerical model, including the initial and boundary conditions, and
the model parameters adopted for the simulations. In \S 3, we present the
results of our hydrodynamical simulations for jets propagating into
stratified ambient media with positive and negative density gradients, and
different values of the density ratio between jet and ambient material, $%
\eta $, and of the Mach number, $M_a$. Finally, in \S 4, we summarize our
results and discuss their possible applications to the observed protostellar
jets.

\section{Description of the Model}

\subsection{Theoretical Grounds}

The basic properties of jets propagating into homogeneous ambient media are
described in BFK and GB93. Similarly to Hardee et al. (1992), we assume an
initial isothermal ambient medium ($T_a=10^4$ $K$) with density (and
pressure) distribution stratified along the jet axis $x$: 
$$
n_a(x)=n_a(x_o)F(x) 
\eqno(1)$$
where 
$$
F(x)=[\alpha (x-x_o)+1]^\beta  
\eqno(2)$$
and $n_a$ is the number density, $x_o$ is the value of $x$ at the jet inlet
(in units of $R_j$, the jet radius), $\alpha $ is a dimensionless
parameter of the model
and the exponent $\beta =\pm 5/3$ for positive and negative density
gradients, respectively. Such profiles were chosen for compatibility with
previous simulations of jets propagating into homogeneous environments.
Also, they are consistent with the observed density distributions of the
clouds which involve protostars (e.g., Fuller and Myers 1992). While a
negative density gradient may represent, for example, the atmosphere that a
jet encounters when it emerges from the protostar and propagates through the
cloud that envelopes the parent source, a positive density gradient may
occur when a fully-emerged jet enters an external cloud (GBB) since the
density of the cloud increases as the jet submerges. In the
simulations, the atmosphere was held steady by an appropriate graviational
potential whose effects on the jet dynamics are neglectable since the jets 
are assumed to be highly supersonic.

A supersonic jet propagating into a stratified ambient gas will develop a
shock pattern at its head whose structure and strength will depend on the
ambient density profile. The impacted ambient material is accelerated by a
forward bow shock whose velocity estimated from momentum flux conservation
in one-dimensional analysis is approximately given by: 
$$
v_{bs}\approx \frac{v_j}{1+(\epsilon \eta )^{-1/2}F(x)^{1/2}} 
\eqno(3)$$
and the jet material is decelerated at the jet shock with a 
velocity $v
_{js}\approx v_j-v_{bs}$. In the equation above $\eta =n_j(x_o)/n_a(x_o)$ is
the ratio between the jet and ambient number densities at the jet inlet ($%
x=x_o)$, and $\epsilon ^{1/2}=R_j/R_h$ is the ratio of jet to head radius.
For a homogeneous ambient medium, for which $F(x)=1$, eq. 3 becomes
identical to eq. 2 of GB93. As long as $\epsilon $ remains constant, eq.
(3) predicts a global acceleration (deceleration) of the head of a $\eta >1$
jet propagating into a negative (positive) density gradient ambient medium 
\footnote{However, the numerical simulations below show
 that $v_{bs}$ often develops an oscillating pattern caused by the combined 
effects
 of $\epsilon$ and the ambient ram pressure ($n_a v_{bs}^2$)
 variations; see \S 4. }. 

For shock velocities $v_s>90$ km s$^{-1}$, the cooling length in the post
shock gas behind the shock is given, in units of the jet radius, by (e.g.
BFK):

$$
q_s\equiv \frac{d_{cool}}{R_j}\approx 8\times 10^{16}R_j^{-1}v_{s,7}^4n^{-1} 
\eqno(4)$$
where $n$ is the preshock number density (we have assumed a number density
of nuclei $\approx $ $n/2$) and $v_{s,7}=v_s/10^7$ cm s$^{-1}$. Using eqs. 2
and 3 above, the cooling parameter for the bow shock is: 
$$
q_{bs}\equiv \frac{d_{cool}}{R_j}\approx 8\times
10^{16}R_j^{-1}v_{j,7}^4n_a(x_o)^{-1}\frac{F(x)^{-1}}{[1+(\epsilon \eta
)^{-1/2}F(x)^{1/2}]^4} 
\eqno(5)$$
and the cooling parameter for the jet shock is: 
$$
q_{js}\approx q_{bs}\epsilon ^{-2}\eta ^{-3}F(x)^3. 
\eqno(6)$$
Eqs. (3, 5, and 6) indicate that, as long as $\epsilon $ remains unaltered
and $\eta >1$, as the jet propagates downstream into an ambient medium with
negative density gradient (decreasing $F(x)$), the gas behind the bow shock
will become increasingly adiabatic ($q_{bs}$ increases), while the
jet shock will become increasingly radiative ($q_{js}$ decreases with
increasing $v_{bs}$ and hence decreasing $v_{js}$). The reverse situation
must hold for jet propagation into an ambient medium with positive density
gradient.

We note that, for example, in an ambient medium with negative density
gradient as $v_{js}$ decreases with increasing $v_{bs}$, it may eventually
become smaller than $90$ kms$^{-1}$. In such a case $q_{js}$ is more
correctly evaluated through (e.g., Falle and Raga 1993)

$$
q_{js}\simeq 0.31(\frac{n_j}{1000\rm{ }cm^{-3}})^{-1}(\frac{v_{js}}{45
\rm{ }km\rm{ }s^{-1}})^{-4.7}(\frac{R_j}{2\times 10^{16\rm{ }}cm})^{-1} 
\eqno(7)$$
where $n_j$ is the jet number density.

The density, $n_{sh}$, of the cold shell that develops in the working
surface from the radiative cooling of the shocked gas can be estimated by
balancing the ram pressure of the ambient medium and the thermal pressure of
the cooled gas: 
$$
n_{sh}KT_{sh}\approx \bar m n_a(x_o)v_j^2\frac{F(x)}{[1+(\epsilon \eta
)^{-1/2}F(x)^{1/2}]^2} 
\eqno(8)$$
The right-hand side gives the ambient ram pressure ($\rho _av_{bs}^2$), $K$
is the Boltzmann constant and $\bar m $ is the mean mass per particle which is
constant in the case of complete ionization. Therefore, for a jet
propagating downstream into a positive-density-gradient-atmosphere, $n_{sh}$
will increase for an $\eta >1$ jet ($\epsilon \sim $constant), as long as [$%
(\epsilon \eta )^{-1/2}F(x)^{1/2}$] remains $<1$. However, as $F(x)$ becomes
very large $n_{sh}$ will tend to an approximately constant value ($%
n_{sh}\approx \bar m n_a(x_o)v_j^2\eta /KT_{sh}\epsilon $).

We note that eqs. 4-8 above apply only in the region close to the axis of
symmetry of the jet, as pressure gradient effects must be non negligible off
the axis. Nonetheless, \S 3 below indicates that the predictions of the
one-dimensional relations above are qualitatively supported by the 3-D
simulations.

\subsection{The Numerical Model}

The hydrodynamics equations are solved using a modified version of the
three-dimensional gridless, Lagrangean smoothed particle hydrodynamics (SPH)
code described in Benz (1990), GB93, GB94 and CMGB. The ambient gas is
represented by a 3-D rectangular box filled with particles. A supersonic jet
of radius $R_j$ is injected continually into the bottom of the box, which
has dimensions up to $\sim 30R_j$ in the x-axis and $12R_j$ in the
transverse directions y and z. All distances are normalized to the jet
radius. The boundaries of the box are assumed to be continuous ($\partial v=0
$) - particles are removed from the system whenever they cross the
boundaries. (In the original code (GB93), the transverse boundaries were
assumed to be periodic. With the present modification to continuous
boundaries, the formation of undesirable reflective waves on the transverse
boundaries, later on the simulations, is prevented.) The SPH particles are
smoothed out in space with a spherically symmetric kernel function of
initial width 0.4 or $0.5$ and 0.2 or $0.25$ $R_j$ for the ambient and jet
particles, respectively.

As in the previous works (GB94, GB94, and CMGB), the jet and the ambient gas
are treated as a single ionized fluid with a ratio of specific heats $\gamma
=5/3$ and an ideal gas equation of state $p=u(\gamma -1)\rho $, where $u$ is
the internal energy per unit mass, $p$ is the thermal pressure, and $\rho
=n\bar m $ is the mass density.

The radiative cooling (due to collisional excitation and de-excitation, and
recombination) is calculated using the cooling function evaluated by Katz
(1989) for a gas of cosmic abundances cooling from $T=10^6$ K. The cooling
is suppressed below $T\approx 10^4$ K when the assumption of completely
ionized flow breaks down and the effects of transfer of ionizing radiation
become important. A time-implicit scheme combined with Newton-Raphson (NR)
iterations is used to evolve the cooling rate. To monitor accuracy, the
maximum change in a timestep is kept less than 30\%. If this condition is
not achieved, or if the system fails to converge in 100 NR iterations, the
timestep is reduced (to $\Delta t/2$) and a solution is once again pursued.
The limitations of the cooling assumptions are discussed in GB93 and GB94.
By not taking into account the effects of nonequilibrium ionization of the
gas or the transfer of ionizing radiation, we are probably underestimating
the cooling rate in some parts of the postshock regions by as much as an
order of magnitude ($e.g.$, Innes, Giddings, \& Falle 1987). However, a
comparison of our results (see also GB93) with multidimensional calculations
for steady flows in a homogeneous ambient medium, which have included a
nonequilibrium time-dependent cooling (SN), show that the essential
dynamical features do not change significantly under the presence of
nonequilibrium ionization effects.

The evolution of the system is parameterized by the dimensionless numbers:
i) $\eta =n_j/n_a$ (the ratio between the input jet and ambient number
densities at $x_o$ (eq. 2); (ii) $M_a=v_j/c_a$ (the initial ambient Mach
number, where the ambient sound speed is given by $c_a=(\gamma kT_a/\bar
m)^{1/2}$, with $T_a$ being the ambient temperature (held constant even in
stratified atmospheres) and $\bar m\simeq 0.5m_H$ the mean mass per particle
for a fully ionized gas of cosmic abundances); (iii) $k_p=p_j/p_a$ (the
input pressure ratio which has been assumed to be initially equal to unity
in all simulations); and (iv) $q_{bs}(0)$ (eq. 5). The stratification of the
ambient medium described by eqs. 1 and 2 adds two new parameters to the
models, $\alpha $ and $\beta $.

Based on typical conditions found in protostellar jets, the parameters of
the simulations were chosen to be the following: density ratio $\eta \simeq
1-10$ ($e.g.$, Mundt, Br\"ugel \& Buhke 1987, Morse et al. 1992, Raga \&
Noriega-Crespo 1993); ambient number density $n_a(0)$ $\simeq 200$ cm$^{-3}$%
; jet velocities $v_j$ $\simeq 200,$ $400$ km s$^{-1}$ (e.g., Reipurth
1989a, Reipurth et al 1992); initial ambient Mach number $M_a=12,$ $24$; jet
radius $R_j$ = $2\times 10^{15}$ cm ($e.g$., Raga 1993). For the parameters
of the stratified ambient medium (eqs. 1 and 2) we assume $\alpha =0-0.5$
and $\beta =\pm 5/3$. We note that the values adopted for $\alpha$ and
$\beta$, correspond to large pressure and density gradients,
which allow us to investigate structural effects on the simulated 
beams without having to make them propagate too many jet radii. In fact, 
large gradients do exist in the ISM through which the jets propagate, 
as we have remarked in \S 1, and the evolution of these gradients 
with time is much slower than the evolution of the jet structures, so 
that the assumption of a nearly steady atmosphere is reasonable. 

The limitations to our SPH simulations are discussed in GB93, GB94 and CMGB,
in particular, we remark that turbulence which is active in these flows
since the Reynolds numbers are very high, $>10^4$ (GB93), is difficult to
study because our initial particle spacing is quite large relative to the
size of eddies that may develop in the flow and the numerical viscosity of
the code may be too dissipative. Thus, for this work we can only consider
the bulk properties of the flow, i.e., over a size scale larger than that of
most of the largest eddies where the internal turbulent motions are averaged
out.

\section{The Simulations}

We have performed runs of radiative cooling, supersonic jets using different
values of $\eta $, $M_a$, and the $\alpha $ and $\beta $ parameters for the
stratified ambient medium. The effects of stratification on adiabatic jets
were also examined. The subsections below present the results of these
calculations and Table 1 summarizes the values of the input 
parameters of our simulations.

\subsection{Effects of ambient media with different ($\alpha ,\beta $)}

Fig. 1 depicts the central density contours and the velocity distribution of
five cooling jets at a time $t=0.85$ (in units of $t_{ja}=R_j/c_a=38.2$ yrs)
evolving through ambient media with: a) and b) positive 
density stratification; 
c) no
stratification; and
d) and e) negative 
density stratification (see Table 1). 
All the jets
have the initial conditions: $\eta =$3, $n_a(x_o)=$200 cm$^{-3}$, $R_j$ = $%
2\times 10^{15}$ cm, $v_j=398$ km s$^{-1}$, $M_a=24$, $q_{bs}(x_o)\approx
8.1 $ (corresponding to an initial bow shock velocity $v_{bs}\approx 252$ km
s$^{-1}$), $q_{js}(x_o)\approx 0.3$ (corresponding to an initial jet shock
velocity $v_{js}\approx 146$ km s$^{-1}$). 
The maximum density reached by
the cold shell that develops at the head of the jet from the cooling of the
shock-heated gas, 
$n_{sh}/n_a(x_o)$,
is in  rough agreement with eq. 8. We clearly
see an acceleration of the jet due to the decrease of the ambient density
and ram pressure ($n_av_{bs}^2$) from Figs. 1a to 1e. Consistent with eq.
(3), the propagation velocity of the beam, $v_{bs}$, increases with
increasing $\eta $.

Figs. 2, 3 and 4 compare the density and velocity evolution of three cases
with $\eta =3$. In Fig. 2, the ambient medium is homogeneous ($\alpha =0$, $%
\beta =0$); in Fig. 3, the ambient medium has positive density gradient ($%
\alpha =0.5$, $\beta =5/3$); and in Fig. 4, it has negative density gradient
($\alpha =0.5$, $\beta =-5/3$). The initial conditions are the same as in
Fig. 1. 

In the homogeneous medium (Fig. 2), we identify the same features detected
in previous works (e.g., BFK, SN, GB93, GB94, CMGB). At the head of the jet,
a dense shell of shocked jet material develops. It becomes Rayleigh-Taylor
(GB93) unstable and breaks into blobs. The density of the shell also
undergoes variations with time due to $global$ $thermal$ $instabilities$ of
the radiative shock (see GB93 for details). 
A thin, low density cocoon containing post-shock jet gas is
deposited around the beam. A shroud of shocked ambient gas envelops both the
beam and the cocoon.

In the increasing density (and pressure) medium (Fig. 3), the cocoon/shroud
is compressed and pushed backwards by the increasing ram pressure of the
ambient medium (see also GBB). An elongated structure (Fig. 3c, d) replaces
the bow shock seen in front of the head of the jet propagating into the
uniform medium of Fig. 2. The increasing pressure and density of the
surrounding medium collimate the beam and promotes the development of the
Kelvin-Helmholtz (K-H) instability (e.g. Birkinshaw 1991), and the inherent
discreteness of the SPH code in the beam excites pinch and helical modes
(Fig. 3c). These modes over-confine the beam, drive internal shocks, and
cause some jet wiggling (see also Figs. 7a, b, and 8 below). It is
interesting to note that this beam structure is similar to the one found for 
$adiabatic$ jets propagating into homogeneous ambient medium (GB93). In the
latter cases, however, the collimation of the beam is a result of the hot,
high pressure, post shock jet gas in the cocoon immediately behind the
working surface.

On average, jets propagating into an ambient medium of increasing density
gradient are decelerated by the increasing ambient ram pressure ($%
n_av_{bs}^2$). However, for the jet of Fig. 3 (see also Figs. 7 and 8
below), we find that the propagation velocity, $v_{bs}$, has an oscillating
pattern caused by competition between the $F(x)$ and $\epsilon (x)$ terms in
eq. 3. Since $F(x)$ increases with distance, $v_{bs}$ initially decreases as
the jet propagates downstream with $\epsilon ^{1/2}\approx 1$. As the jet
head is compressed by the ambient medium, however, the increase of $\epsilon 
$ more than compensates for the decrease of $F(x)$ and the jet head
accelerates as it narrows. Eventually, the further decrease of $F(x)$ again
dominates and the working surface decelerates again. Table 2 shows the
velocity variations in the head of the jet of Fig. 3 as a function of the
time and the position of the bow shock ($x_{bs}-x_o$) (see also GBB). The
varying bow shock and jet shock velocity ($v_{js}\approx v_j-v_{bs}$) may
cause the ambient ram pressure ($n_av_{bs}^2$; Table 2) and the radiative
cooling distances behind both shocks, $q_{bs}$ and $q_{js,}$ to fluctuate
(eqs. 5 and 6), and the working surface to become $thermally$ unstable
(GB93). On average, however, the bow shock becomes more radiative and the
jet shock more adiabatic as the jet propagates downstream, in qualitative
agreement with eqs. 5, 6 and 7. Following the oscillating pattern, the
maximum density in the shell at the head $n_{sh}/n_a(x_o)$ varies from: $%
\approx $ 847 ($t/t_{ja}=$0.85), to 3550 ($t/t_{ja}=$1.25), 651 ($t/t_{ja}=$%
1.65), and 741 ($t/t_{ja}=$1.95), in Figs. 3a, b, c, and d, respectively.
The internal shocks of the beam have peak separations $\sim 1-2R_j$ and
travel downstream with a similar variable velocity as the working surface.
The shocks closer to the jet head have their speed closer to $v_{bs}$.

In the decreasing density (and pressure) medium (Fig. 4), the jet retains a
well-developed bow shock and cavity structure. The beam shows a slight
decollimation close to the head ($\epsilon ^{1/2}\approx 0.7$ in Figs.
4c, and d; see also Fig. 9). $On$ $average,$ it propagates downstream with
increasing velocity, ($v_{bs}\approx $ 285, 325, 353, and 376 km $s^{-1}$,
in Figs. 4a, b, c, and d, respectively), due to the declining ambient ram
pressure, in rough agreement with eqs. 1, 2, and 3. The cold shell at the
head is thin 
and without
fragmentation. No internal shocks are formed along the beam. For an ambient
medium with negative density gradient (increasing $\eta $), the cooling
length behind the bow shock ($q_{bs}$) increases as the jet propagates
downstream (eqs. 4, 5): $q_{bs}$ increases from 8.1 at t=0 to $\sim $3958 at
t = 1.45 $t_{ja}$, so that the bow shock becomes more adiabatic as the beam
propagates. By contrast, $q_{js}$  initially decreases from 0.3 at
t=0 to $\sim $0.1 at t=0.25 $t_{ja}$, in agreement with eq. 6, then
increases to $\sim $149 at t=1.45 $t_{ja}$ (eq. 7), so that the jet shock
becomes less radiative. 

For comparison, Figs. 5a, b, and c show the mid-plane radiative emissivity (%
$L=n^2\Lambda (t)$) contours of the jets of the Figs. 2, 3, and 4,
respectively, when their working surfaces are at a position $x\simeq 25R_j$.
 In
the decreasing density ambient medium (Fig.5c), the jet is very faint with a
maximum emissivity which is $\sim $1.3$\times $10$^{-2}$ of that of the jet
in the homogeneous ambient medium (Fig. 5a). The latter, in turn, has a
maximum emissivity which is $\sim $1.6$\times $10$^{-2}$ of that of the jet
in the increasing density ambient medium (Fig. 5b). The time evolution of
the radiative emissivity of the head of the jets of Figs. 2, 3, and 4 in the
axis ($y=z=0$) is depicted in Fig. 6. As the jet in the decreasing density
ambient medium ($\alpha =0.5$, $\beta =-5/3$) propagates downstream the
emissivity (which is generally dominated by the jet shocked material) decays
and the shock becomes mostly adiabatic. For the jet propagating into the
homogeneous ambient medium ($\alpha =0$, $\beta =0$) the varying pattern of
the emissivity is correlated with the density oscillations of the
cold shell discussed above. The emissivity of the jet propagating into the
increasing density medium ($\alpha =0.5$, $\beta =5/3$)  (which is initially
dominated by the bow shock and after t$\simeq 1.25$ $t_{ja}$ by the internal
knot closest to the head) increases with time, as expected. 

\subsection{Jets with different density ratio relative to the ambient medium 
$\eta $}

Fig. 7 depicts the velocity distribution of the central slice of cooling jets
with different $\eta $ propagating into an $\alpha =0.5$ ambient medium with
positive ($\beta =5/3$) density (and pressure) stratification (see the
corresponding density maps in Fig. 2 of GBB). The jets have for $\eta $=1: $%
q_{bs}(x_o)\approx 3.1$; $q_{js}(x_o)\approx 3.1$; $\eta $=3: $%
q_{bs}(x_o)\approx 8.1$, $q_{js}(x_o)\approx 0.3$; and $\eta $=10: $%
q_{bs}(x_o)\approx 16.7$, $q_{js}(x_o)\approx 0.008$, and other initial
conditions the same as in previous runs (Table 1). 
The $\eta =1$ jet is the most
evolved, reaching the end of the computation domain at $t=1.85$ $t_{ja}$,
while the $\eta =3$ jet reaches the end at $t=1.65$ $t_{ja}$ and the $\eta
=10$ jet at $t=1.25$ $t_{ja}$. 
The maximum density of the shells at the jet
heads are $n_{sh}/n_a(x_o)\approx $ 211, 638, and $2540$, respectively. 
Jets
with larger $\eta $ are less affected by the increasing ambient density and
pressure gradients, and show less focusing by the compressing ambient
medium. Consistently, the development of the K-H modes (see also Fig. 3),
which cause the beam pinching and wiggling, are stronger on small-$\eta $
jets. The internal oblique shocks driven by beam pinching have a typical
separation of a few jet radii, in agreement with the observed internal knots
of Young Stellar Object jets (e.g., MRR). Like
the $\eta =3$ jet (see Fig. 3), both the $\eta =1$ and $\eta =10$ jets also
show oscillatory propagation velocities due to the competing effects of the
increasing ambient density and the ratio $\epsilon ^{1/2}$ (eq. 3). Tables 3
and 4 show the velocity variations in the heads of the $\eta =1$ and $10$
jets. In the case of the $\eta =10$ jet, the velocity oscillations are
smoother, and the smaller velocity variation in this case is not enough to
cause significant fluctuations on the ambient ram pressure ($n_av_{bs}^2$),
which decreases with increasing distance. The internal shocks of Figs. 7a,
and b travel downstream under a similar variable velocity pattern as the
working surface.

Fig. 8 shows that later ($t=2.05$ $t_{ja}$), the knots close to the head of
the $\eta $=1 jet break and $separate,$ but this effect may have been
amplified by the nearby continuitive boundary.

Fig. 9 depicts three cooling jets with different $\eta $ propagating into an 
$\alpha =0.5$ ambient medium with negative ($\beta =-5/3$) density (and
pressure) stratification and the same initial conditions as Fig. 7. 
As in Fig. 4, in all cases the cocoon becomes broad and relaxed. The cold
shell is thin and its density is low (see Fig. 9), 
in qualitative agreement with eq. 7. The beam
decollimation is significantly larger for the smallest $\eta $ jet.

\subsection{Jets with different M$_a=v_j/c_a$}

Fig. 10 depicts an $\eta =3$ cooling jet propagating into an $\alpha =0.5$
ambient medium with positive ($\beta =5/3$) density (and pressure)
stratification and $M_a$=12 (or $v_j$=199 km s$^{-1}$), which can be
compared with the jet of Fig. 3 for which $M_a$=24. The initial values of $%
q_{bs}(x_o)\approx 0.5$ (eq. 5) and $q_{js}(x_o);\approx 1.9\times 10^{-2}$
(eq. 7) for $\eta =3$ (Table 1)
imply that both shocks initially cool very fast and the
jet shock is nearly isothermal (e.g. GB93). 
The jet reaches the end of the
domain at $x=20R_j$ at $t$= 3.85 $t_{ja}$ (Fig. 10c). This snapshot can be
compared with Fig. 3c where the $M_a$=24 jet has propagated about the same
distance (at t = 1.65 $t_{ja}$). The smaller $M_a$ jet seems to be less
affected by the focusing effect of the ambient medium with increasing
pressure, but the interface with the cocoon is more affected by turbulent
entrainment (CMGB).

Fig. 11 depicts a jet propagating into an $%
\alpha =0.5$ ambient medium with negative ($\beta = -5/3$) density (and
pressure) stratification (compare this $M_a$=12 jet with 
the $M_a$=24 case in
Fig. 4). 
The beam decollimation close to the head caused
by the decreasing ambient pressure is slightly larger in the smaller, more
evolved $M_a$ jet. The maximum density in the cold shell at the jet head is
also smaller in this latter case.  

\subsection{Adiabatic versus Radiative Cooling Jets}

Fig. 12 depicts the central density contours and velocity distribution of an
adiabatic and a radiative cooling jet (as in Fig. 3), both with $\eta $= 3
and $M_a$ = 24 propagating into an $\alpha =0.5$ ambient medium with
positive ($\beta =5/3$) density (and pressure) stratification at $t=$1.45 $%
t_{ja}$. The ambient medium compresses the cooling jet more than the
adiabatic jet, and the thinner, lower-pressure cocoon around the cooling jet
is less able to preserve its structure. The propagation velocities are very
similar until $t\approx $1.45 $t_{ja}$, when, due to the larger focusing ($%
>\epsilon ^{1/2}$) of the head, the radiative jet propagates about twice as
fast. The maximum density behind the shock is $n_{sh}/n_a(x_o)\approx $650
for the cooling jet and 6 for the adiabatic jet.

Fig. 13 depicts the density contours and velocity distribution of an
adiabatic and a radiative cooling jet (also in Fig. 4), both with $\eta $= 3
and $M_a$ = 24 propagating into an $\alpha =0.5$ ambient medium with
negative ($\beta =5/3$) density (and pressure) stratification, at $t=$0.85 $%
t_{ja}$ when they reach x$\approx 20R_j$. The propagation velocity $v_{bs}$,
the decreasing ambient ram pressure, and the head decollimation are very
similar in both jets, although slightly larger in the adiabatic case$.$ The
density behind the shock is: $n_{sh}/n_a(x_o)\approx $32 for the cooling jet
and 6 for the adiabatic jet.

\section{Discussion and Conclusions}

In this work, we have presented fully 3-D hydrodynamical simulations of
initially overdense, radiative cooling jets propagating into stratified
isothermal ambient media with a power-law density and pressure distribution
(eq. 1). We also have investigated the behaviour of adiabatic jets
propagating into similar stratified environments. As in our previous work
(GB93, GB94, CMGB94, GBB), the simplified treatment of the radiative cooling
of the gas prevents us from performing a detailed comparison of our models
with the radiation from observed protostellar jets. In particular the
density contour maps, while describing reasonably well the expected gas
distribution do not necessarily correspond to the observed emission line
images. Nonetheless, with the help of those maps, we can delineate some
basic structural characteristics and dynamics, and compare with observed
protostellar jets.

Our results show that jets propagating into regions of an ambient medium
with negative density (pressure) gradient develop a broad and relaxed cocoon
and suffer beam decollimation at the head. Their propagation velocity
increases, on average, as they propagate downstream due to the drop of the
ambient density and ram pressure. The beam decollimation is larger for the
small-$\eta $ jets as their momentum flux is smaller (Figs. 4 and 7). No
internal knots are formed in these cases and the weak radiative shocks at
the head provide little radiation. This suggests a possible explanation for
the $invisible$ portions of the observed outflows, as we have proposed in
GBB95. For example, the observed gaps close to the source of Haro 6-5B (or
FS Tau B), AS 353, or 1548 C27 jets (e.g. MRR; Hartigan, Mundt, and Stocke
1986; Mundt et al. 1984, 1987), or the gaps between the beam and the distant
bow shock (e.g. Haro 6-5B, ) could be a product of such an effect (see
also GBB).

Jets propagating into an ambient medium with increasing density (pressure)
have their cocoon/shroud compressed and pushed backwards by the ambient ram
pressure and the beam is highly collimated. The compressing medium promotes
beam focusing and wiggling, and the formation of traveling internal shocks
via beam pinching. The bow shock like structure at the head disappears. Jets
with larger $\eta $ are less focused by the compressing medium, as a
consequence of their larger momentum flux. We found that within $\sim 30R_j$
a $M_a=24$ ($v_j\approx $398 km s$^{-1}$) jet with $\eta \leq 3$,
propagating into a positive density gradient medium wiggles and develops
regularly-spaced traveling knots close to the head (see also GBB).

Protostellar jets are generally well collimated and knotty, and often show
one or more bow shock features at the head (e.g. Reipurth 1989a; Hartigan,
Raymond, and Meaburn 1990). A number of authors (e.g. Hartigan and Raymond
1992, Kofman and Raga 1992, Raga and Kofman 1992, SN; GB94) have shown that
the internal traveling bright knots may be produced by small-period
variations in the velocity of jet ejection. This mechanism favors the
formation of knots closer to the driving source which become fainter and
disappear at larger distances, as for example, in jet systems like HH34 and
HH111 (e.g. Reipurth 1989b). On the other hand, BFK and GB93 have shown that
due to radiative cooling, K-H instabilities were less effective at driving
internal shocks in beams propagating into a $homogeneous$ ambient medium. In
the present work, however, we find that propagation of a radiative cooling
jet into an ambient medium of increasing density (pressure) may drive the
formation of internal knots close to the jet head. This seems to be the
case, for example, for the Haro 6-5B northern and HH 24G jets (Mundt et al.
1990, MRR), which present an elongated wiggling and knotty structure far
from the source, very similar to those of Figs. 3 and 7 (see GBB for a
detailed discussion).

Large amplitude side-to-side wiggles and knots are also observed in the
HH83, HH84, HH85, HH110 (e.g., Reipurth 1989a, b), and HH30 jets (Mundt et
al 1990, MRR), all of which are immersed in regions. The HH30
jet, for example, belongs to the HLTau/HH30 region which shows five
outflows and a number of reflection nebulae associated with YSO's. As for
Haro 6-5B jet, we suggest that the observed collimated, wiggling, and knotty
structure of these jets could be due to K-H instabilities driven
by a compressing medium of increasing density (as in Figs. 3 and 7). An
alternative, recently discussed by Raga, Cant\'o,  and Biro (1993) is that the
jet is ejected from a precessing source into a homogeneous ambient medium.

Also, our results indicate that increasing density environments 
destroy the bow shock at the head. Some authors have demonstrated that bow
shocks can be formed in jets with continuous ejection (e.g., Norman et al.
1982; BFK; GB93) or with multiple outflow episodes of long period (e.g.,
Hartigan and Raymond 1992, Raga and Kofman 1992; SN; GB94) and can keep
their structure stable in portions of the jet where the ambient medium is
more $homogeneous$ (GB93). However, there are cases which show no evidence
of a terminal bow shock feature (e.g. some sources of the MRR sample), which
may match some of our results (Figs. 3, 7, and 8). There are other cases
(e.g. Haro 6-5B (MRR), and HH30 jets (L\'opez et al. 1995) in which the
elongated beam is separated by a long gap from the distant bow shock. The
bow shock in these cases could be a relic of an earlier outflow episode, or
the place where the jet impacts a dense portion of the ambient medium after
a long expansion through a decreasing density (pressure) medium (see GBB).

Although, on average, the jets propagating into an increasing (decreasing)
density environment are decelerated (accelerated) by the increasing
(decreasing) ram pressure of the ambient medium, we find that their
propagation velocities $v_{bs}$ have an oscillating pattern due to the
competing effects between the varying density ratio $\eta $ and the ratio of
jet to head radius $\epsilon ^{1/2}=R_j/R_h$. The internal traveling shocks
that develop in jets propagating into positive density gradient environments
(Figs. 3, 7, and 8) also follow a variable velocity pattern, with their
speeds getting closer to $v_{bs}$ as they approach the jet head. This is in
qualitative agreement with recent measurements of the tangential velocity of
the knots of Haro 6-5B (Eisl\"ofel 1993). Variations of the shock velocity
must be accompanied by variations in the intensity of emission from cooling
regions behind the shocks. Thus, further detections of intensity and knot
proper motion variability in other systems would provide a possible check of
our predictions.

In the comparison of jets with different initial Mach numbers $M_a$
propagating into ambient portions of increasing density, we find that (Figs.
3 and 10) the smaller $M_a$ jet ($M_a=12)$ seems to be less affected by the
focusing effect of the ambient medium with increasing pressure. Due to its
smaller pressure cocoon, the interface between the cocoon and the smaller $%
M_a$ jet is more affected by lateral turbulent entrainment (e.g. Raga,
Cabrit and Cant\'o 1994; Gouveia Dal Pino 1995). In a recent investigation
of radiative cooling jets propagating into $homogeneous$ ambient media
(CMGB), we found that the lateral entrainment of ambient gas through
turbulent mixing at the contact discontinuity is relevant only for low Mach
number $(M_j=M_a$ $\eta ^{1/2}\leq 3),$ low density ratio ($\eta \leq 3)$
flows which, in general, are not appropriate for protostellar jets for which 
$10<M_j<40$ and $\eta $ $>1$. Entrainment of ambient gas seems to be mostly
through the bow shock in protostellar jets (CMGB). These results have ruled
out models which proposed that the molecular outflows associated with
protostellar jets are produced predominantly through turbulent entrainment
(e.g., Stahler 1993). In the present study, however, we find that jets with $%
M_j=21$ (or $M_a=12$ and $\eta =3$) in an ambient medium of increasing
density suffer considerable entrainment along the beam. This result
suggests that even in high Mach number jets, turbulent entrainment may play
an important role in driving molecular outflows $if$ the ambient medium has
regions of varying density (and pressure). We note that the large
line broadening (FWHM$\simeq $ 100-150 km s$^{-1}$) detected along portions
of some of the jets of HLTau/HH30 complex (Mundt et al. 1990) is an
indication that turbulent entrainment may be relevant in those regions, as
suggested by Mundt et al.

We also performed short-time runs of adiabatic jets into stratified
environments in order to compare their structure with that of radiative
cooling jets (Figs 12 and 13). In general, the radiative cooling jet
propagating into an ambient medium of positive density gradient is more
collimated by the compressing medium than the adiabatic jet. On the other
hand, the adiabatic jet propagating into an ambient medium of negative
density gradient suffers more head decollimation. Both effects are due to
the fact that the adiabatic jet has a higher pressure cocoon, which is
better able to preserve the beam structure in the compressing environment,
and to accelerate the ambient gas where its density is
decreasing.

Finally, as in our previous investigations (GB93, GB94, CMGB94, GBB), in
this study we have assumed a history-independent optically thin radiative
cooling function to comput the losses of a fully ionized flow. By not
following the history-dependent effects of nonequilibrium ionization of the
gas or the transfer of ionizing radiation, we possibly underestimated the
cooling rate in some parts of the postshock regions by as much as an order
of magnitude (e.g., Innes, Giddings \& Falle 1987). The inclusion of a
detailed cooling evaluation, as in the one-dimensional calculations of HR,
would require a substantial increase in computer power. Although future work
should take into account those effects, we expect that the gross dynamical
features obtained in the present analysis will not change. This expectation
is supported by the comparison of our results (see GB93, GB94) with the
calculations of SN (see also their recent calculations for steady flows;
 Stone \& Norman 1993b, 1994b), which included a nonequilibrium time-dependent
cooling.

\vspace{1in}

E.M.G.D.P. would like to acknowledge the very fruitful discussions with
Willy Benz, Reinhard Mundt, and Tom Ray. 
The authors are also indebted to the referee for his relevant remarks. 
The simulations were performed on
the Workstations HP apollo 9000/720 and Dec 3000/600 AXP of the Plasma
Astrophysics Group of the Department of Astronomy of IAG/USP, whose purchase
was made possible by the Brazilian agency FAPESP. This work was partially
supported by a CNPq grant.

\newpage

\begin{center}
{\bf TABLE 1} Values of the input parameters of our models.

\begin{tabular}{||c|c|c|c|c|c|c|c|c|c||} \hline
$Figure$ & $\alpha$ & $\beta$ & $\eta$ & $M_a$ & $q_{bs}(x_o)$ 
& $q_{js}(x_o)$
\\ \hline 
  1a & 0.5 &  5/3  &  3  & 24 & 8.1 & 0.3  \\
  1b & 0.1 &  5/3  &  3  & 24 & 8.1 & 0.3  \\
  1c & 0   &  0    &  3  & 24 & 8.1 & 0.3  \\
  1d & 0.1 &  -5/3 &  3  & 24 & 8.1 & 0.3  \\
  1e & 0.5 &  -5/3 &  3  & 24 & 8.1 & 0.3  \\
  2, 5a & 0   &  0    &  3  & 24 & 8.1 & 0.3  \\
  3, 5b & 0.5 &  5/3  &  3  & 24 & 8.1 & 0.3  \\
  4, 5c & 0.5 &  -5/3 &  3  & 24 & 8.1 & 0.3  \\
  7a, 8 & 0.5 &  5/3  &  1  & 24 & 3.1 & 3.1  \\
  7b & 0.5 &  5/3     &  3  & 24 & 8.1 & 0.3  \\
  7c & 0.5 &  5/3     &  10  & 24 & 16.7 & 0.008  \\
  9a & 0.5 &  -5/3    &  1  & 24 & 3.1 & 3.1  \\
  9b & 0.5 &  -5/3    &  3  & 24 & 8.1 & 0.3  \\
  9c & 0.5 &  -5/3    &  10  & 24 & 16.7 & 0.008  \\
  10 & 0.5 &  5/3     &  3  & 12 & 0.5 & 0.019  \\
  11 & 0.5 &  -5/3    &  3  & 12 & 0.5 & 0.019  \\
  12 & 0.5 &  5/3     &  3  & 24 & 0   & 0  \\
  13 & 0.5 &  -5/3    &  3  & 24 & 0   & 0  \\
\hline
\end{tabular}

\end{center}

\newpage

\begin{center}
{\bf TABLE 2} Velocity oscillations in the $\eta =3$ jet head of Fig. 3 (and
Fig. 7b) propagating into an environment of increasing density ($\alpha =0.5$%
, $\beta =5/3$). 

\begin{tabular}{||c|c|c|c|c|c|c|c|c|c||} \hline
$t/t_{ja}$  &  $(x_{bs} - x_o)/R_j$  & $(\epsilon)^{1/2}$  & $v_{bs}$ 
$(km s^{-1})$ & 
$n_a v_{bs}^{2\dag}$ \\ \hline 
  0  &  0  & 1  & 250 & $2.3 \times 10^{2}$  \\ 
  0.45&  5.3  &   1.2  & 200 & $1.2 \times 10^3$ \\
  0.85 & 8.8 & 1.4 & 144 & $1.2 \times 10^3$ \\
  1.25 & 12 & 1.6 & 130 & $1.7
\times 10^{3}$ \\
  1.65 &  18.7 & 3.3 & 260 & $1.2 \times 10^4$  \\
  1.95
& 23 & 3.2 & 225 & $1.2 \times 10^4$ \\  
\hline
\end{tabular}

\end{center}

\hspace*{1cm}{\footnotesize $^{\dag }$ n$_a$ v$_{bs}^{2}$%
  \ is in units of n$_a$(x$_o$) c$_a^2$ = 5.5 $\times 10^{14}
$ cm$^{-1}$s$^{-2}$ .}

\begin{center}
{\bf TABLE 3} Velocity oscillations in the head of the $\eta =1$ jet of Fig.
7a propagating into an environment of increasing density ($%
\alpha =0.5$, $\beta =5/3$). 

\begin{tabular}{||c|c|c|c|c|c|c|c|c|c||} \hline
$t/t_{ja}$  &  $(x_{bs} - x_o)/R_j$  & $(\epsilon)^{1/2}$  & $v_{bs}$ 
$(km s^{-1})$ & 
$n_a v_{bs}^{2\dag}$ \\ \hline 
  0  &  0  & 1  & 199 & $1.4 \times 10^{2}$ \\ 
  0.85  &  7.7  &   1.3  & 125 & $7.9 \times 10^{2}$ \\
 1.25 & 13.1 & 6 & 224 & $5.3 \times 10^3$  \\
 1.85 &  18.9 & 5 & 126 & $2.9 \times 10^3$ \\
\hline
\end{tabular}
\end{center}

\newpage
\begin{center}
{\bf TABLE 4} Velocity oscillations in the head of the $\eta =10$ jet of
Fig. 7c propagating into an environment of increasing density ($%
\alpha =0.5$, $\beta =5/3$). \\ 

\begin{tabular}{||c|c|c|c|c|c|c|c|c|c||} \hline
$t/t_{ja}$  &  $(x_{bs} - x_o)/R_j$  & $(\epsilon)^{1/2}$  & $v_{bs}$ 
$(km s^{-1})$ & 
$n_a v_{bs}^{2\dag}$ \\ \hline 
 0  &  0  & 1  & 302 & $3.3 \times 10^2$ \\ 
 0.45  &  7.5  &   1.1  & 232 & $2.6 \times 10^3$  \\
 0.65 & 10.5 & 1.7 & 249 & $4.8 \times 10^3$ \\
 1.05 &  15.9 & 1.7 & 212  & $6.3 \times 10^3$ \\
 1.25 & 19 & 2 & 259  & $1.2 \times 10^4$  \\
\hline
\end{tabular}
\end{center}

\newpage


\newpage

\begin{center}
{\bf {Figure Captions} \\ }
\end{center}

\setlength{\baselineskip}{24pt}

Figure 1. Mid-plane density contours and velocity distribution of five
radiative cooling jets at t= 0.85 $R_j/c_a$ ($=38.2$ yr) evolving through
different stratified ambient media with: a) $\alpha =0.5,$ $\beta =5/3$, b) $%
\alpha =0.1,$ $\beta =5/3$, c) $\alpha =0,$ $\beta =0$, d) $\alpha =0.1,$ $%
\beta =-5/3$, and e) $\alpha =0.5,$ $\beta =-5/3$. The initial parameters
are $\eta =3,$ $n_a=200$ cm$^{-3}$, $R_j=2\times 10^{15}$ cm, $M_a=24$, and $%
v_j=398$ km/s. The z and x coordinates are in units of $R_j$. 
The contour
lines are separated by a factor of $1.3$ and the density scale covers the
range from $\simeq 0.02$ up to $363/n_a$.
The maximum density reached by
the cold shell developed at the head of the jet is
$n_{sh}/n_a(x_o)\approx $: a) 363, b) 294, c) 197, d)
69, and e) 32. 

Figure 2. Mid-plane density contour and velocity distribution evolution
of a radiative cooling jet propagating into a homogeneous ambient medium ($%
\alpha =0$, $\beta =0$). The initial conditions are the same as in Fig. 1 ($%
\eta =3,$ $n_a=200$ cm$^{-3}$, $R_j=2\times 10^{15}$ cm, $M_a=24$, and $%
v_j=398$ km/s). The contour lines are separated by a factor of $1.3$ and the
density scale covers the range from $\simeq 0.16$ up to $400/n_a$. The
times  (in units of $t_{ja}=R_j/c_a=38.2$ yr) and jet head positions
depicted are: a) 0.70 $t_{ja}$ and 10.6 $R_j$; b) 1.05 $t_{ja}$ and 17.6 $R_j
$;  c) 1.40 $t_{ja}$ and 24.6 $R_j$; and d) 1.75 $t_{ja}$ and 31.6 $R_j$.
The maximum density
of the cold shell at the head, 
$n_{sh}/n_a(x_o),$ is: $\approx $ a) 195, b) 400, c) 238, and
d) 272. 

Figure 3. Mid-plane density contour and velocity distribution evolution of a
radiative cooling jet propagating into an ambient medium with positive
density (and pressure) gradient ($\alpha =0.5$, $\beta =5/3$). The initial
conditions are the same as in  Fig. 2. The contour lines are separated by a
factor of $1.2$ and the density scale covers the range from $\simeq 0.01$ up
to $3550/n_a$. The times  and jet head positions are: a) 0.85 $t_{ja}$
and 8.8 $R_j$; b) 1.25 $t_{ja}$ and 12 $R_j$; c) 1.65 $t_{ja}$ and 18.7 $R_j$%
; and d) 1.96 $t_{ja}$ and 23 $R_j$.

Figure 4. Mid-plane density contour and velocity distribution evolution of a
radiative cooling jet propagating into an ambient medium with negative
density (and pressure) gradient ($\alpha =0.5$, $\beta =-5/3$). The initial
conditions are the same as in Fig. 2. The contour lines are separated by a
factor of $1.2$ and the density scale covers the range from $\simeq 0.01$ up
to $29/n_a$. The times and jet head positions are: a) 0.25 $t_{ja}$ and
4.3 $R_j$; b) 0.65 $t_{ja}$ and 12.3 $R_j$; c) 1.05 $t_{ja}$ and 20.8 $R_j$;
and d) 1.45 $t_{ja}$ and 30 $R_j$.
The maximum density constrast in the thin shell 
at the head is 
$n_{sh}/n_a(x_o)\approx $13,
23, 31, and 18, 
in Figs. 4a, b, c, and d, respectively.

Figure 5. Mid-plane radiative emissivity contour of the jets of Figs. 2, 3,
and 4 when they reach $x\simeq 25$ $R_j$. a) jet in the homogeneous ambient
medium ($\alpha =0$, $\beta =0$) at t =1.4 $t_{ja}$; b) jet in the
increasing density ambient medium ($\alpha =0.5$, $\beta =5/3$) at t =1.96 $%
t_{ja}$; c)  jet in the decreasing density ambient medium ($\alpha =0.5$, $%
\beta =-5/3$) at t =1.25 $t_{ja}$. 
The emissivity is in units of 
$3.8 \times 10^{-19}$ erg cm$^{-3}$ s$^{-1}$.
The contour
lines are separated by a factor of $1.2$ and the emissivity scale covers the
range from $\simeq 0.01$ up to $3.3 \times 10^5$ in code units.

Figure 6. Time evolution of the axial radiative emissivity of the  head of
the jets of Figs. 2 ($\alpha =0$, $\beta =0$), 3 ($\alpha =0.5$, $\beta =5/3$%
), and 4 ($\alpha =0.5$, $\beta =-5/3$).  
The emissivity is in units of 
$3.8 \times 10^{-19}$ erg cm$^{-3}$ s$^{-1}$.

Figure 7. Velocity distribution of the mid-plane of cooling jets with
different $\eta $ propagating into an $\alpha =0.5$ ambient medium with
positive ($\beta =5/3$) density (and pressure) stratification: a) $\eta $=1:
b) $\eta $=3: and c) $\eta $=10. The corresponding density contours are
shown in Fig. 2 of GBB. The other initial conditions are the same of the
previous figures. The contour lines are separated by a factor of $1.3$ and
the density scale covers the range from $\simeq 0.01$ up to $2540/n_a$. The $%
\eta =1$ reaches the end of the computation domain at  $t=1.85$ $t_{ja}$;
the $\eta =3$ at $t=1.65$ $t_{ja}$ and the $\eta =10$ jet at $t=1.25$ $t_{ja}
$. 

Figure 8. Mid-plane density contour and the corresponding velocity
distribution of the $\eta $=1 jet of Fig. 7 at $t=2.05$ $t_{ja}$. The
contour lines are separated by a factor of $1.3$ and the density scale
covers the range from $\simeq 0.05$ up to $205/n_a$. The knots close to the
head seem to break and $separate$ apart.

Figure 9. Mid-plane density contour and the corresponding velocity
distribution of cooling jets with different $\eta $ propagating into an $%
\alpha =0.5$ ambient medium with negative ($\beta =-5/3$) density (and
pressure) stratification and the same initial conditions of Fig. 5: a) $\eta 
$=1: b) $\eta $=3: and c) $\eta $=10.The jets reach the end of the domain,
at $x\approx 20R_{j\rm{,}}$ at $t=:$a) 1.05 $t_{ja}$ ($\eta =1$); b) 0.95 $%
t_{ja}$ ($\eta =3$); and c) 0.85 $t_{ja}$ ($\eta =10$). The contour lines
are separated by a factor of $1.3$ and the density scale covers the range
from $\simeq 0.01$ up to $58/n_a$.
The maximum density in the cold shell at the head is
$n_{sh}/n_a(x_o)\approx $4.4,
33, 
and 58, respectively. 

Figure 10. Mid-plane density contour and velocity distribution evolution of
an $\eta =3$ cooling jet propagating into an $\alpha =0.5$ ambient medium
with positive ($\beta =5/3$) density (and pressure) stratification and $M_a$%
=12 (or $v_j$=199 km s$^{-1}$). The other initial conditions are the same of
Fig. 3, except that $q_{bs}(x_o)\approx 0.5$ (eq. 5) and $%
q_{js}(x_o);\approx 1.9\times 10^{-2}$. The contour lines are separated by a
factor of $1.3$ and the density scale covers the range from $\simeq 0.01$ up
to $284/n_a$. The times depicted are: a) 1.15 $t_{ja}$; b) 2.45 $t_{ja}$; and
c) 3.85 $t_{ja}$.

Figure 11. Mid-plane density contour and velocity distribution evolution of
an $\eta =3$ cooling jet propagating into an $\alpha =0.5$ ambient medium
with negative ($\beta =5/3$) density (and pressure) stratification and $M_a$%
=12 (or $v_j$=199 km s$^{-1}$). The other initial conditions are the same of
Fig. 4 except that $q_{bs}(x_o)\approx 0.5$ (eq. 5) and $q_{js}(x_o);\approx
1.9\times 10^{-2}$. The contour lines are separated by a factor of $1.3$ and
the density scale covers the range from $\simeq 0.01$ up to $14/n_a$. The
times depicted are: a) 0.65 $t_{ja}$; b) 1.25 $t_{ja}$; and c) 1.85 $t_{ja}$.
The jet reaches the end of the domain at $x\approx 20R_j$. 
The maximum density in the cold shell is
$n_{sh}/n_a(x_o)\approx $14, 11, and 6 in
the successive frames of the Figure.

Figure 12. Central density contour and velocity distribution of an adiabatic
(a) and a radiative cooling (b) jet, both with $\eta $= 3 and $M_a$ = 24
propagating into an $\alpha =0.5$ ambient medium with positive ($\beta =5/3$%
) density (and pressure) stratification at $t=$1.45 $t_{ja}$. The contour
lines are separated by a factor of $1.3$ and the density scale covers the
range from $\simeq 0.01$ up to $650/n_a$.

Figure 13. The density contours and velocity distribution of an adiabatic
(a) and a radiative cooling (b) jet, both with $\eta $= 3 and $M_a$ = 24
propagating into an $\alpha =0.5$ ambient medium with negative ($\beta =-5/3$%
) density (and pressure) stratification at $t=$0.85 $t_{ja}$ when they reach
x$\approx 20R_j$. The contour lines are separated by a factor of $1.3$ and
the density scale covers the range from $\simeq 0.01$ up to $32/n_a$.

\end{document}